\documentclass[aps,prl,preprintnumbers,twocolumn,superscriptaddress,nofootinbib,floatfix,10pt]{revtex4}
\usepackage{amsfonts,amssymb,stmaryrd,latexsym,amsmath,braket}
\usepackage{graphicx,subfigure}
\usepackage{times}
\usepackage{slashed}
\usepackage{braket}
\usepackage{comment}
\usepackage[utf8]{inputenc}
\newcommand{\beginsupplement}{%
        \setcounter{table}{0}
        \renewcommand{\thetable}{S\arabic{table}}%
        \setcounter{figure}{0}
        \renewcommand{\thefigure}{S\arabic{figure}}%
        \setcounter{equation}{0}
        \renewcommand{\theequation}{S\arabic{equation}}        
     }
\begin{document}
        
\title{Convergence of Eigenvector Continuation}

\author{Avik Sarkar}
\email{sarkarav@msu.edu}
\affiliation{Facility for Rare Isotope Beams and Department of Physics and Astronomy,
Michigan State University, East Lansing, MI 48824, USA}

\author{Dean Lee}
\email{leed@frib.msu.edu}
\affiliation{Facility for Rare Isotope Beams and Department of Physics and Astronomy,
Michigan State University, East Lansing, MI 48824, USA}
        
        \begin{abstract}
        Eigenvector continuation is a computational method that finds the extremal eigenvalues and eigenvectors of a Hamiltonian matrix with one or more control parameters.  It does this by projection onto a subspace of eigenvectors corresponding to selected training values of the control parameters.  The method has proven to be very efficient and accurate for interpolating and extrapolating eigenvectors.  However, almost nothing is known about how the method converges, and its rapid convergence properties have remained mysterious.  In this letter we present the first study of the convergence of eigenvector continuation. In order to perform the mathematical analysis, we introduce a new variant of eigenvector continuation that we call vector continuation.  We first prove that eigenvector continuation and vector continuation have identical convergence properties and then analyze the convergence of vector continuation.  Our analysis shows that, in general, eigenvector continuation converges more rapidly than perturbation theory.  The faster convergence is achieved by eliminating a phenomenon that we call differential folding, the interference between non-orthogonal vectors appearing at different orders in perturbation theory.  From our analysis we can predict how eigenvector continuation converges both inside and outside the radius of convergence of perturbation theory.  While eigenvector continuation is a non-perturbative method, we show that its rate of convergence can be deduced from power series expansions of the eigenvectors.  Our results also yield new insights into the nature of divergences in perturbation theory.

        \end{abstract}
\maketitle

        Eigenvector continuation (EC) is a variational method that finds
the extremal eigenvalues and eigenvectors of a Hamiltonian matrix that depends
on one or more control parameters \cite{Frame:2017fah}.  The method consists of projecting
the Hamiltonian onto a subspace of basis vectors corresponding to eigenvectors at some chosen
training values of the control parameters. It has been used to extend quantum Monte Carlo
methods to problems with strong sign oscillations \cite{Frame:2019jsw}, as a fast emulator for
quantum many-body systems \cite{Konig:2019adq,Ekstrom:2019lss}, and as a resummation method for perturbation theory \cite{Demol:2019yjt}.  Eigenvector continuation is well suited for studying the connections between microscopic nuclear forces and nuclear structure, a topic that has generated much recent interest \cite{Ekstrom:2015rta,Elhatisari:2016owd,Lapoux:2016exf,Piarulli:2017dwd,Lu:2018bat,Binder:2018pgl,Soma:2019bso,Gandolfi:2020pbj,Tews:2020hgp}.  All of these applications would be greatly enhanced with a better fundamental understanding of the convergence of the method.  For that purpose, in this letter we present the first study of the convergence properties of eigenvector continuation. 

Let us consider a one-parameter family of Hamiltonian matrices $H(c) = H_0 + cH_1$, where both $H_0$ and $H_1$ are finite-dimensional Hermitian matrices.  The extension to the multi-parameter case will be discussed after our discussion of the one-parameter case. For the given Hamiltonian family $H(c)$, we are interested in finding the ground state eigenvector $\ket{v(c_t)}$ and eigenvalue $E(c_t)$ for some target parameter value $c = c_t$. The objective of eigenvector continuation is to approximate $\ket{v(c_t)}$ as a linear combination of ground state eigenvectors $\ket{v(c_0)}, \cdots, \ket{v(c_N)}$ at training points $c=c_0,\cdots , c_N$. In eigenvector continuation the best linear combination of training-point  vectors is chosen by minimizing the expectation value of $H(c_t)$. As noted in Ref.~\cite{Frame:2017fah}, eigenvector continuation can also be extended to excited states by including excited state eigenvectors at the training points.  However we will focus on ground state calculations in this analysis.

It is convenient to introduce a variant of eigenvector continuation which we call vector continuation (VC). Like eigenvector continuation, in vector continuation we approximate $\ket{v(c_t)}$ as a linear combination of vectors $\ket{v(c_0)}, \cdots, \ket{v(c_N)}$ at training points
$c=c_0,\cdots , c_N$. The difference is that in vector continuation
we construct the best approximation  by projecting $\ket{v(c_t)}$ onto the subspace spanned by the training point vectors. This is a simpler process than the variational calculation used in eigenvector continuation.  Since it requires knowledge of the target eigenvector, vector continuation should be viewed as a diagnostic tool rather than a method for determining $\ket{v(c_t)}$.  We will show that eigenvector continuation and vector continuation have identical convergence properties, and so it suffices to understand the convergence properties of vector continuation. As the name suggests, vector continuation can also be generalized to any smooth
vector path $\ket{v(c)}$ without reference to Hamiltonian matrices or eigenvectors. 

We want to understand the asymptotic convergence of eigenvector and vector continuation at large orders.  To do this we consider a sequence of training points $c_0,\cdots , c_N$ with a well-defined limit point $c_{\rm lim}$ for large $N$ that is some distance away from the target point $c_t$.  Although we are taking the limit of large $N$, the values of $N$ we probe are always vastly smaller than the number of dimensions of our linear space.  Our choice of training points provides a good definition for the convergence properties at large orders.  It can be viewed as the worst possible conditions for convergence, where all of our training points are clustered in one area but we need to extrapolate to a target point located somewhere else.  In future work we plan to discuss the extension to the case where the training points are clustering around more than one accumulation point.  When the training points are spread apart and not clustered, the convergence is generally much faster, especially if the training points surround the target point from all sides.  However, the convergence for this more general case is highly dependent on the positions of the training points, and it is difficult to make a clean definition of asymptotic convergence.  Nevertheless, in the Supplemental Materials we show that convergence for the multi-parameter case is much faster if there exists a smooth curve connecting some subset of the training points to the target point.

 Without loss of generality we can redefine $c$ so that our limit point corresponds to $c=0$.  In the limit where the training points accumulate around\ $c=0$, we can replace our training vectors $\ket{v(c_0)}, \cdots, \ket{v(c_N)}$ with the derivatives of $\ket{v(c)}$ at $c=0$, which we write as $\ket{v^{(0)}(0)}, \cdots, \ket{v^{(N)}(0)}$.  These derivative vectors approximately span the same $N+1$-dimensional subspace as our original training vectors.  In the following we write $\ket{v(c_t)}^{\rm EC}_N$ for the order-$N$ eigenvector continuation approximation to $\ket{v(c_t)}$, and we write $\ket{v(c_t)}^{\rm VC}_N$ for the order-$N$ vector continuation approximation to $\ket{v(c_t)}$. We use the same set of training vectors $\ket{v^{(0)}(0)}, \cdots, \ket{v^{(N)}(0)}$ for eigenvector continuation and vector continuation. 
 
 %As we vary $c$, we can study the vector path described by the ground state eigenvector of $H(c)$ with the help of eigenvector continuation. However, for a given $c_t$, in order to perform EC, we only need to know the training eigenvectors, which are all near $c=0$. Neither EC nor VC needs the knowledge of the path of the eigenvector as we vary $c$. As such, in both cases we do not make reference to the vector path described by eigenvector of $H(c)$. All we need is existence of such a path as we vary $c$. If we were to choose our training points differently and space them apart from themselves, then we can think about the effect of choosing different paths, in which our training points lie, on the convergence of EC. However, this choice of the path of the training points makes sense when we have more than one parameter in our Hamiltonian, and we discuss this further in the supplemental section.

Starting from the derivative vectors $\ket{v^{(0)}(0)}, \cdots, \ket{v^{(N)}(0)}$, we use Gram-Schmidt orthogonalization to define a sequence of orthonormal vectors $\ket{w^{(0)}(0)}, \cdots, \ket{w^{(N)}(0)}$. Here we make the assumption that the derivative vectors are linearly independent, which is generally true for all of the practical problems we encounter. With this orthonormal basis, we can write $\ket{v(c_t)}^{\rm VC}_N$ as 
\begin{equation}
                \ket{v(c_t)}^{\rm VC}_N = \sum_{n=0}^{N}\braket{w^{(n)}(0)|v(c_t)}\ket{w^{(n)}(0)}.
                \label{VC}
        \end{equation}
Using the same orthonormal basis, we can also write $\ket{v(c_t)}^{\rm EC}_N$ as
\begin{equation}
                \ket{v(c_t)}^{\rm EC}_N = \sum_{n=0}^{N}a(c_t,n,N)\ket{w^{(n)}(0)},
                \label{EC}
        \end{equation}
where the coefficients $a(c_t,n,N)$ are found by minimizing the expectation value of $H(c_t)$.

We now consider perturbation theory (PT) around the point $c=0$.  If $z_0$ is the nearest branch point to $c=0$, then the series expansion 
        \begin{equation}
                \ket{v(c_t)} = \sum_{n=0}^{\infty}\ket{v^{(n)}(0)}\frac{c_t^n}{n!}
                \label{PT}
        \end{equation}
will converge for $|c_t|<|z_0|$ and diverge for $|c_t|>|z_0|$.
We define $\ket{v(c_t)}^{\rm PT}_N$ as the partial series truncated at order $n=N$, 
\begin{equation}
                \ket{v(c_t)}^{\rm PT}_N = \sum_{n=0}^{N}\ket{v^{(n)}(0)}\frac{c_t^n}{n!}.
                \label{PTN}
        \end{equation}
In this analysis we have assumed that the radius of convergence is greater than zero.  This follows from the fact that $H(c)$ is a finite-dimensional Hermitian matrix for all real $c$. In a forthcoming publication we will discuss the extension to infinite-dimensional systems and the interesting case where the radius of convergence is zero.

\begin{figure}
               % \centering
                \includegraphics[width=8.4cm]{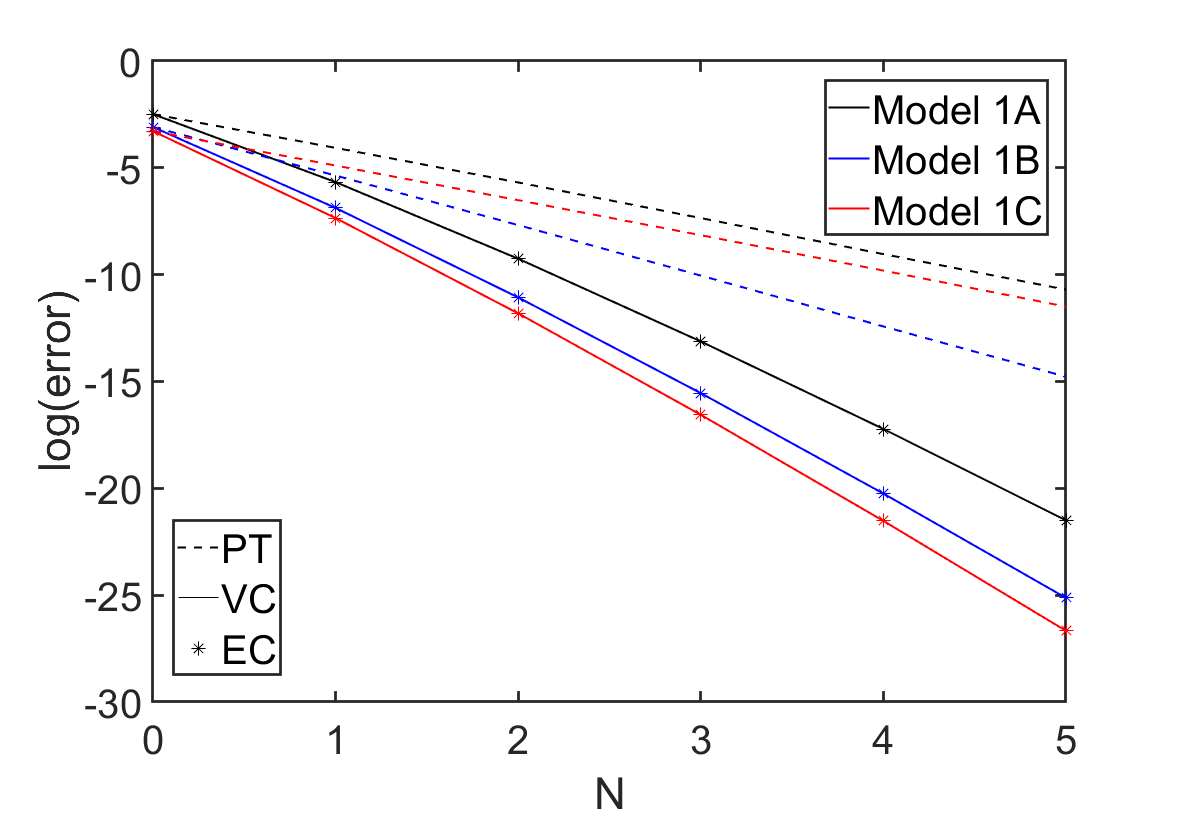}\\
                \caption{(Color online) Logarithm of the error versus order
$N$ for eigenvector continuation (asterisks), vector continuation (solid
lines), and perturbation theory (dashed lines).  The three different colors
(black, blue and red) correspond with Models 1A, 1B, and 1C respectively.}
                \label{error_plot_1}
        \end{figure}

We will quantify the error of these three approximations to $\ket{v(c_t)}$ by computing the norm of the residual vector as a function of $N$. From the norm of the residual vector versus $N$, we can determine whether the approximations are converging or not. In Fig.~\ref{error_plot_1} we plot the logarithm of the error 
versus order $N$ at fixed $c_t$. The results are shown for eigenvector continuation (asterisks), vector continuation
(solid lines), and perturbation theory (dashed lines).  The three different colors correspond to three different examples, which we call Models 1A, 1B, and 1C, each with a linear space of 800 dimensions. For each example we see that eigenvector and vector continuation converge more rapidly than perturbation theory.  Furthermore, eigenvector and vector continuation have nearly identical errors at each order.  The Hamiltonians for the three models are given in the Supplemental Materials.  There is nothing special about these matrix models, and we find similar results to these in all matrix examples where perturbation theory is convergent.    

Let us now prove that eigenvector continuation and vector continuation indeed have identical convergence properties.  We first consider vector continuation at order $N$.  Let $V^N(0)$ be the subspace spanned by 
$\ket{w^{(0)}(0)}, \cdots \ket{w^{(N)}(0)}$, and let $V^N_\perp(0)$ be the orthogonal complement.
As one can see from Eq.~(\ref{VC}), there is no error at all in the coefficients of $\ket{w^{(0)}(0)}, \cdots \ket{w^{(N)}(0)}$.  The residual vector for $\ket{v(c_t)}^{\rm VC}_N$ lies entirely in $V^N_\perp(0)$.

We now consider eigenvector continuation at order $N$.  In this case we project $H(c_t)$ onto $V^N(0)$ and find the 
resulting ground state.  In essence, we have turned off all 
matrix elements of $H(c_t)$ that involve 
vectors in $V^N_\perp(0)$.  Let us now turn on these matrix elements as a perturbation.  When these matrix elements are turned back on, we will get a first-order correction to the wave function from transition matrix elements connecting $V^N(0)$ with $V^N_\perp(0)$.  This will produce a correction 
to the wave function that lies in $V^N_\perp(0)$.  
On the other hand, the corrections to the coefficients of $\ket{w^{(0)}(0)}, \cdots \ket{w^{(N)}(0)}$ will appear at second order in perturbation theory, since this involves transitions from $V^N(0)$ to $V^N_\perp(0)$ and then returning back from $V^N_\perp(0)$ to $V^N(0)$.  If the norm of the residual vector for eigenvector continuation is $O(\epsilon)$, then eigenvector continuation and vector continuation will differ at $O(\epsilon^2)$. This proves that eigenvector continuation and vector continuation have identical convergence properties in the limit of large $N$.

Let us now consider the norm of the ``last term'' corresponding to $n=N$ in Eqs.~(\ref{VC}), (\ref{EC}), and (\ref{PTN}),
\begin{align}         
L^{\rm VC}_N(c_t) & = \left| \braket{w^{(N)}(0)|v(c_t)} \right|, \label{LVC} \\
L^{\rm EC}_N(c_t) & = \left| a(c_t,N,N) \right|, \label{LEC} \\
L^{\rm PT}_N(c_t) & = \left\| \ket{v^{(N)}(0)}\frac{c_t^N}{N!} \label{LPT} \right\|.
\end{align}
Combining Eq.~(\ref{LVC}) with Eq.~(\ref{PT}), we have
        \begin{equation}
                 L^{\rm VC}_N(c_t) = \left| \sum_{n=N}^{\infty} \braket{w^{(N)}(0)|{v^{(n)}(0)}}\frac{c_t^n}{n!} \right|.
                \label{VC-PT}
        \end{equation}
In this series expression, we will refer to the partial series up to $n=N$ as the leading order (LO) approximation.  We will call the partial series up to $n=N+1$ the next-to-leading order (NLO) approximation, and so on.
The N$^k$LO approximation is therefore 
        \begin{equation}
                 L^{{\rm VC},{\rm N}^k{\rm LO}}_N(c_t) = \left| \sum_{n=N}^{N+k} \braket{w^{(N)}(0)|{v^{(n)}(0)}}\frac{c_t^n}{n!}
\right|,
                \label{VC-PT2}
        \end{equation}
and at leading order we have
\begin{equation}
L^{\rm VC, LO}_N(c_t) = \left| \braket{w^{(N)}(0)|{v^{(N)}(0)}}\frac{c_t^N}{N!}
\right|. \label{LVC_LO}
\end{equation}
By comparing Eq.~(\ref{LVC_LO}) with Eq.~(\ref{LPT}), we can understand why vector continuation is converging more rapidly than perturbation theory. In general, $\left|\braket{w^{(N)}(0)|{v^{(N)}(0)}}\right|$ is smaller than the norm of $\ket{v^{(N)}(0)}$ because  $\ket{v^{(N)}(0)}$ is not orthogonal to the lower derivative vectors $\ket{v^{(0)}(0)}, \cdots \ket{v^{(N-1)}(0)}$.  Perturbation theory must deal with constructive and destructive interference between non-orthogonal vectors at different orders, a phenomenon that we call differential folding.  Differential folding can be a very large effect, and it is the reason why perturbation theory converges more slowly than vector and eigenvector continuation.

In order to study the convergence properties systematically, let us define the convergence ratio obtained by taking two widely separated orders $N'$ and $N$, with $N > N'$, and computing the quantities 
\begin{align}
\mu^{\rm VC}(c_t) &= \left|L^{\rm VC}_N(c_t) /L^{\rm VC}_{N'}(c_t) \right|^{1/(N-N')}, \label{VCNN'} \\
\mu^{\rm EC}(c_t) &= \left|L^{\rm EC}_N(c_t) /L^{\rm EC}_{N'}(c_t) \right|^{1/(N-N')},
\label{ECNN'} \\
\mu^{\rm PT}(c_t) &= \left|L^{\rm PT}_N(c_t) /L^{\rm PT}_{N'}(c_t) \right|^{1/(N-N')}.
\label{PTNN'}
\end{align}
For notational convenience, we omit writing the explicit dependence on $N$ and $N'$.  When it is clear from the context, we also omit writing the dependence on $c_t$.  These definitions are motivated from the ratio test of convergence for a series. Intuitively, $\mu$ is ratio at which consecutive terms in the series converge (or diverge) asymptotically. We note that these convergence ratio functions will have cusps where the numerator vanishes and divergences where the denominator vanishes.  Fortunately these special points occur at only a few isolated values of $c_t$, and the functions in Eq.~(\ref{VCNN'}), (\ref{ECNN'}), and (\ref{PTNN'}) provide a useful picture of the convergence properties of the three methods. We can eliminate cusps or divergences at any particular value of $c_t$ by changing $N$ or $N'$.

%The convergence ratio for eigenvector continuation, $\mu^{\rm EC}$, can be calculated directly using Eq.~(\ref{LEC}). In our analysis here, we are using the orthogonalized derivatives vectors $\ket{w^{(0)}(0)}, \cdots, \ket{w^{(N)}(0)}$ as our training vectors for eigenvector continuation.

   \begin{figure}
                \centering
                \includegraphics[width=8.4cm]{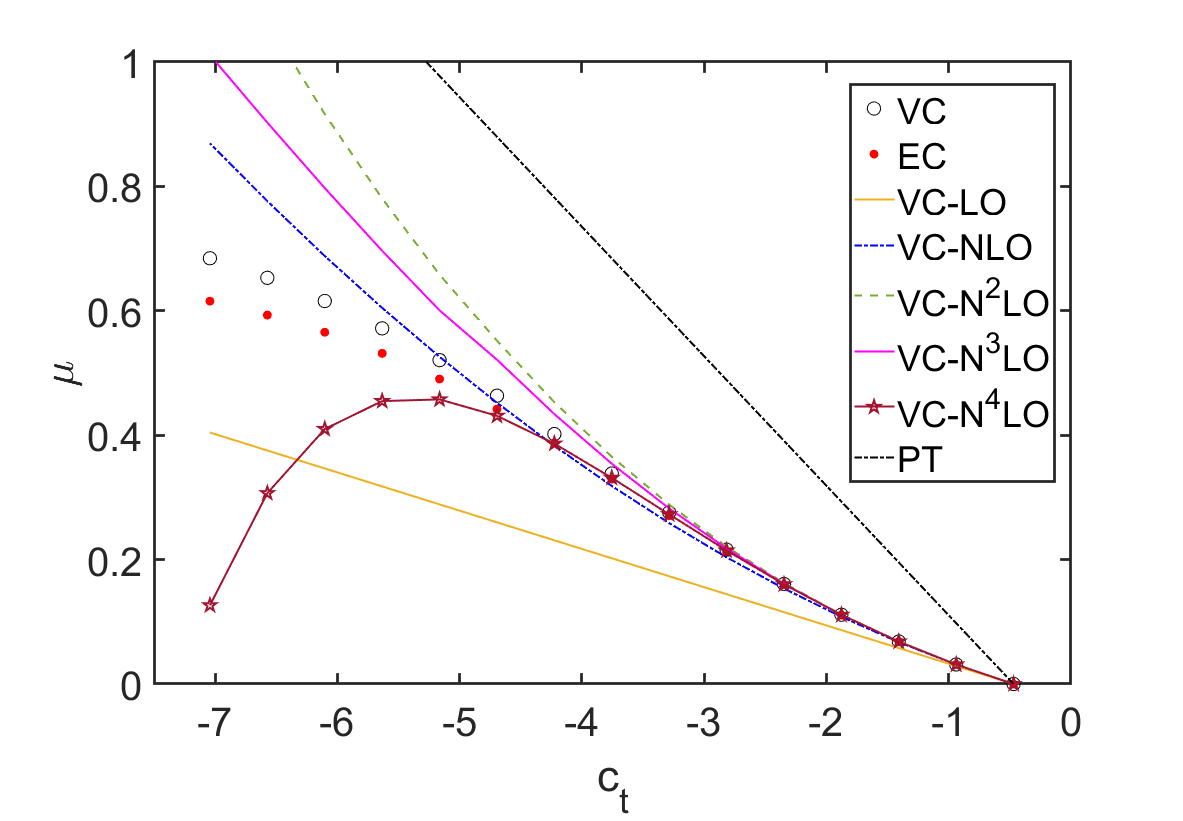}\\
                \caption{(Color online) Comparison of the convergence ratios $\mu^{\rm VC}(c_t)$,
$\mu^{\rm EC}(c_t)$, and $\mu^{\rm PT}(c_t)$ for Model 2 with $N = 10$ and $N' = 0$. The training vectors for all cases are evaluated on the weak-coupling BCS side at $c = -0.4695$.  The unitary limit value corresponds to $c = -3.957$.} 
                \label{EC_PT_VC}
        \end{figure}

In our next example, Model 2, we consider a system of two-component fermions with attractive zero-range interactions in three dimensions.  At weak coupling the many-body system forms a Bardeen-Cooper-Schrieffer (BCS) superfluid, while at strong coupling it behaves as a Bose-Einstein condensate (BEC) \cite{Leggett:1980pro,Nozieres:1985JLTP}. In between the BCS and BEC regimes, there is a smooth crossover region that contains a scale-invariant point called the unitary limit where the scattering length diverges and the two-body system has a zero energy resonance. There have been numerous experimental studies of BCS-BEC crossover and the unitary limit using trapped ultracold Fermi gases of alkali atoms
\cite{OHara:2002,Gupta:2002,Regal:2003,Ku:2012a}. Here we use eigenvector continuation to study the crossover transition for two spin-up and two spin-down fermions in an $L = 4$ periodic cubic lattice as detailed in Ref.~\cite{Bour:2011ad}. The Hamiltonian for this system corresponds to a linear space with 262144 dimensions and is described in the Supplemental Materials.  Our control parameter $c$ corresponds to the product of the particle mass $m$ and interaction coupling $C$, as measured in dimensionless lattice units.

The parameter value $c=-3.957$ corresponds with the unitary limit, with larger negative values corresponding to the strong-coupling BEC phase, and smaller negative values corresponding to the weak-coupling BCS phase.  For this example the point $c = 0$ corresponds to a non-interacting system with special symmetries and degeneracies, and so we choose the training vectors at a more general point on the weak-coupling BCS side at $c = -0.4695$.  In Fig.~\ref{EC_PT_VC} we show the convergence ratios $\mu^{\rm VC}$, $\mu^{\rm EC}$, and $\mu^{\rm PT}$ versus $c_t$ for $N = 10$ and $N' = 0$.  We see that $\mu^{\rm VC}$ and $\mu^{\rm EC}$ remain well below $\mu^{\rm PT}$, indicating the faster convergence of vector and eigenvector continuation compared to perturbation theory.  As we cross into the strong-coupling BEC side, perturbation theory diverges, as indicated by the convergence ratio $\mu^{\rm PT}$ exceeding $1$.  However vector and eigenvector continuation both converge even at very strong coupling far on BEC side, as indicated by $\mu^{\rm VC}$ and $\mu^{\rm EC}$ both remaining well below $1$.  Our findings point to a very intriguing future area of study where the superfluid many-body wave function can perhaps be reconstructed by variational methods throughout the entire BEC-BCS crossover region.  We note that the vector and eigenvector continuation results are in close agreement with each other, with only a slight difference when the convergence is slower.  In the same figure, we plot the LO, NLO, N$^2$LO, N$^3$LO, and N$^4$LO approximations to $\mu^{\rm VC}$ as defined in Eq.~(\ref{VC-PT2}).  We see that the expansion of $\mu^{\rm VC}$ converges for $c_t$ with a radius of convergence coinciding with that of perturbation theory.  
                        
\begin{figure}
      \centering
           \includegraphics[width=8.4cm]{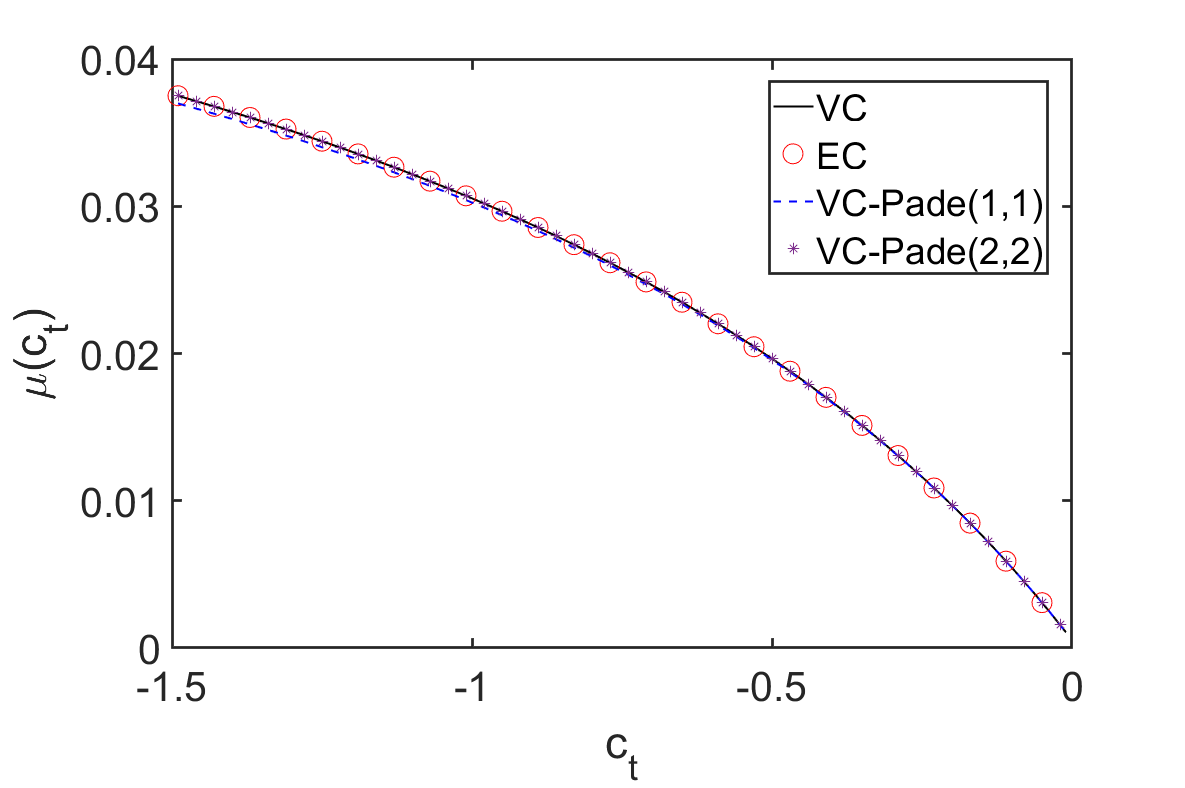}\\
           \caption{(Color online) Plots of the convergence ratios $\mu^{\rm VC}$, $\mu^{\rm EC}$, and the (1,1) and (2,2) Pad{\'e} approximations to $\mu^{\rm VC}$ versus $c_t$ for Model 3 with $N = 20$ and $N' = 0$.}
                \label{PT_VC_Pade}
        \end{figure}

Outside the radius of convergence of perturbation theory, we can still estimate the convergence ratio using extrapolation methods. If there are no branch points nearby, then the convergence ratio function can be extrapolated using standard methods such as Pad{\'e} approximants \cite{Demol:2019yjt} or conformal mapping \cite{Kompaniets:2016lmy,VanHoucke:2020}. 
 To illustrate this, we consider another example called Model 3 where we can control the location and sharpness of the avoided level crossings of eigenvalues that cause the breakdown of perturbation theory.  Our Model 3 Hamiltonian matrix resides in a linear space with $500$ dimensions and more details are given in the Supplemental Materials.  By design, the closest branch point to $c=0$ occurs very close to the real axis, near the point $c=0.84$. In Fig.~\ref{PT_VC_Pade} we plot $\mu^{\rm VC}$ and $\mu^{\rm EC}$ for $N = 20$ and $N' = 0$, and negative $c_t$, extending beyond the radius of convergence of perturbation theory of Model 3. We also show the (1,1) and (2,2) Pad{\'e} approximations to $\mu^{\rm VC}$. We see that the Pad{\'e} approximations describe the shape of $\mu^{\rm VC}$ quite well since there are no nearby branch points. 
 
If there is a branch point nearby, such as we have for Model 3 near $c=0.84$, the slope of the convergence ratio function will rise more quickly than predicted by Pad{\'e} approximants or conformal mapping.  This is because at the branch point, the Riemann surface of the ground state eigenvector is entwined with the Riemann surface of the first excited state eigenvector.  If the branch point is very close to the real axis, then we have an avoided level crossing or Landau-Zener transition where the wave functions of the ground state and first excited state interchange as we pass by the branch point. 

We can therefore predict the rise of $\mu^{\rm
VC}$ and $\mu^{\rm
EC}$ from the fall of $\mu_1^{\rm
VC}$ and $\mu_1^{\rm
EC}$ for the first excited state.  We define 
$\mu_1^{\rm VC}$ and $\mu_1^{\rm
EC}$ in the same manner as $\mu^{\rm
VC}$ and $\mu^{\rm
EC}$, except that we replace the target ground state $\ket{v(c_t)}$ with the first excited state $\ket{v_1(c_t)}$, while using the same orthonormal basis states $\ket{w^{(n)}(0)}$ associated with the ground state at $c=0$.   For the eigenvector continuation approximation of the first excited state, $\ket{v_1(c_t)}^{\rm EC}_N$, we use a subspace that includes derivatives of the ground state $\ket{v^{(0)}(0)} \cdots \ket{v^{(N)}(0)}$ and also derivatives of the first excited state $\ket{v_1^{(0)}(0)} \cdots \ket{v_1^{(N)}(0)}$.

\begin{figure}
          \centering
                \includegraphics[width=8.4cm]{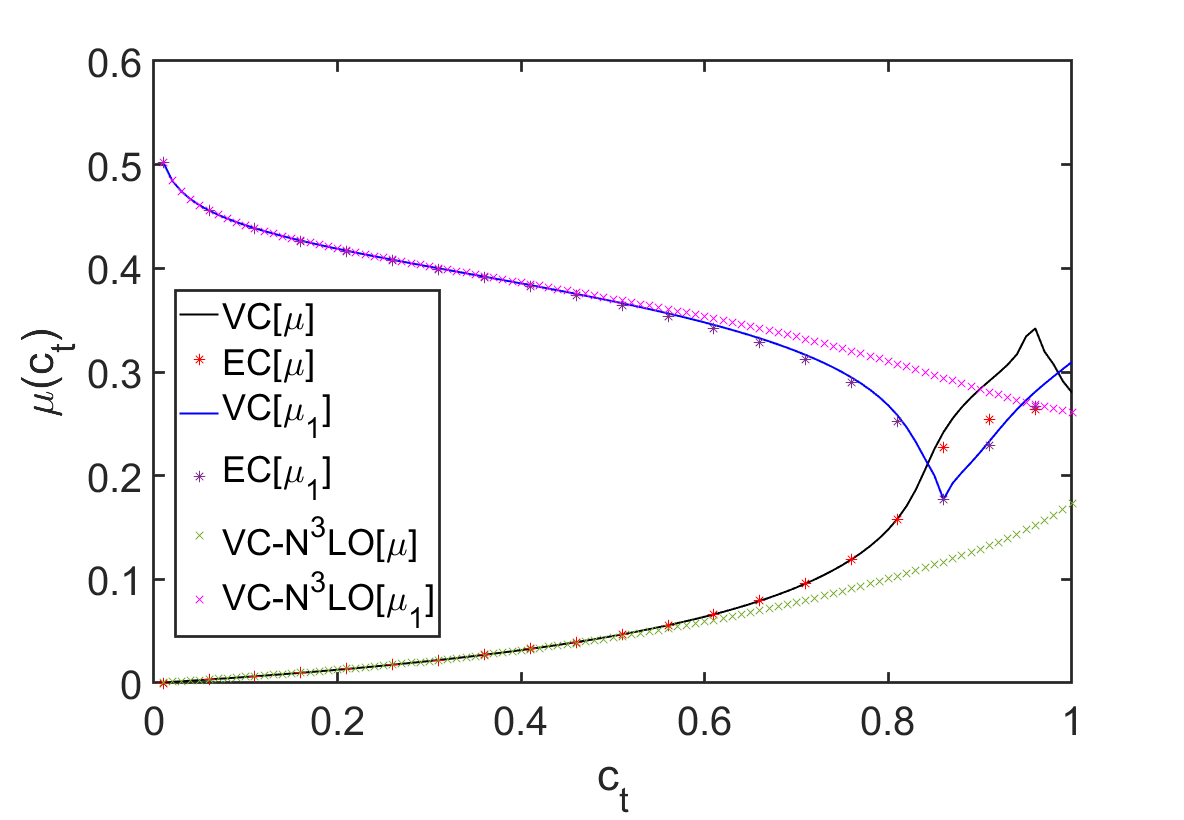}\\
                \caption{(Color online) Plots of the convergence ratios $\mu^{\rm VC}$,
$\mu_1^{\rm VC}$, $\mu^{\rm EC}$,
$\mu_1^{\rm EC}$, and the N$^3$LO
approximations to $\mu^{\rm VC}$ and
$\mu_1^{\rm VC}$ versus $c_t$ for Model 3 with $N = 20$ and $N' = 0$.}
                \label{Fig:excited}
 \end{figure}

In Fig.~\ref{Fig:excited} we show $\mu^{\rm VC}$,
$\mu_1^{\rm VC}$, $\mu^{\rm EC}$,
$\mu_1^{\rm EC}$, and the  N$^3$LO
approximations to $\mu^{\rm VC}$ and
$\mu_1^{\rm VC}$ for $N = 20$ and $N' = 0$.    
We note the approximate vertical and horizontal reflection symmetries near the branch point.  For $c_t < 0.84$ the increase in the ground-state convergence ratio mirrors the decrease in the excited-state convergence ratio. Also the increase in the ground-state convergence ratio for $c_t > 0.84$ mirrors the decrease in the excited-state convergence ratio for $c_t < 0.84$.  

We observe that for this case, a great deal of information about the convergence of vector and eigenvector continuation can be predicted from series expansions around $c=0$.  Near the branch point we know that $\mu^{\rm VC}$, and therefore also $\mu^{\rm EC}$, crosses the midpoint of the gap between N$^k$LO approximations to $\mu^{\rm VC}$ and
$\mu_1^{\rm VC}$ for any $k$.  The N$^k$LO approximations to $\mu^{\rm VC}$ and
$\mu_1^{\rm VC}$ are calculated entirely from perturbation theory at $c=0$.  Also, the location of the nearby branch point can itself be deduced from the convergence radius of the series expansion.  While there
are limits to how far we can go in $c_t$ with these convergence ratio predictions, it is clear that we can predict the convergence ratios both inside and outside the radius of convergence from the derivatives of the eigenvectors near $c=0$. It is quite intriguing that this information can be used to predict the non-perturbative convergence of vector and eigenvector continuation. 

We can now discuss the extension of eigenvector continuation to the case with $D>1$ parameters. The multi-parameter case is exactly equivalent to the one parameter case if we work with directional derivatives $(\vec{c}_t-\vec{c})\cdot\vec{\nabla}$ acting upon $\ket{v(\vec{c})}$.  Here $\vec{c}$ is the limit point of the training data and $\vec{c}_t$ is the target point.  But we now consider the more difficult problem of convergence at all target points at some fixed distance from $\vec{c}$.  If we want to construct the $k^{\rm th}$ directional derivative for any $\vec{c}_t$, we will need all $(k+D-1)!/[k!(D-1)!]$ partial derivatives at order $k$.  Summing over all $k$ from $0$ to $N$ gives $(N+D)!/(N!D!)$.  Hence if we want to achieve the same error as $N+1$ training vectors in the one parameter case, we need $(N+D)!/(N!D!)$ training vectors for the $D$ parameter case.  In the Supplemental Materials we show that this performance analysis is consistent with our numerical results for $D=2$ and $D=3$.

In this letter we have presented the first study of the convergence of eigenvector continuation. 
We found that the series expansion of the wave function exhibits an effect called differential folding, the interference among non-orthogonal terms at different orders.  Both vector and eigenvector continuation avoid this problem.  As a result, they converge faster than perturbation theory and do not diverge for any value of the control parameter.  While most studies of the divergence of perturbation theory focus on series expansions of energy eigenvalues and other observables \cite{Bender:1969si,Parwani:2000rr,Hamprecht:2003vr,Argyres:2012ka,Cherman:2014ofa,Kompaniets:2016lmy,VanHoucke:2020}, our results provide new insights into these divergences as arising from large non-orthogonal terms in the series expansion of the wave function. In our analysis we were able to predict how eigenvector continuation converges outside the radius of convergence of perturbation theory.  All existing and future applications of eigenvector continuation will benefit from this new fundamental understanding of the convergence of the method. 

%\section*{Acknowledgement}
{\it We are grateful for discussions with J. Bonitati, T. Duguet, N. Elkies, G. Given, C. Hicks, S. König, N. Li, B.-N. Lu, and J. Watkins. We acknowledge partial financial support
from the U.S. Department of Energy (DE-SC0018638).   The computational
resources were provided by Michigan State University, RWTH Aachen University, the Oak Ridge Leadership Computing Facility, and the J\"{u}lich Supercomputing Centre.}

\bibliography{References}

\begin{thebibliography}{28}
\expandafter\ifx\csname natexlab\endcsname\relax\def\natexlab#1{#1}\fi
\expandafter\ifx\csname bibnamefont\endcsname\relax
  \def\bibnamefont#1{#1}\fi
\expandafter\ifx\csname bibfnamefont\endcsname\relax
  \def\bibfnamefont#1{#1}\fi
\expandafter\ifx\csname citenamefont\endcsname\relax
  \def\citenamefont#1{#1}\fi
\expandafter\ifx\csname url\endcsname\relax
  \def\url#1{\texttt{#1}}\fi
\expandafter\ifx\csname urlprefix\endcsname\relax\def\urlprefix{URL }\fi
\providecommand{\bibinfo}[2]{#2}
\providecommand{\eprint}[2][]{\url{#2}}

\bibitem[{\citenamefont{Frame et~al.}(2018)\citenamefont{Frame, He, Ipsen, Lee,
  Lee, and Rrapaj}}]{Frame:2017fah}
\bibinfo{author}{\bibfnamefont{D.}~\bibnamefont{Frame}},
  \bibinfo{author}{\bibfnamefont{R.}~\bibnamefont{He}},
  \bibinfo{author}{\bibfnamefont{I.}~\bibnamefont{Ipsen}},
  \bibinfo{author}{\bibfnamefont{D.}~\bibnamefont{Lee}},
  \bibinfo{author}{\bibfnamefont{D.}~\bibnamefont{Lee}}, \bibnamefont{and}
  \bibinfo{author}{\bibfnamefont{E.}~\bibnamefont{Rrapaj}},
  \bibinfo{journal}{Phys. Rev. Lett.} \textbf{\bibinfo{volume}{121}},
  \bibinfo{pages}{032501} (\bibinfo{year}{2018}), \eprint{1711.07090}.

\bibitem[{\citenamefont{Frame}(2019)}]{Frame:2019jsw}
\bibinfo{author}{\bibfnamefont{D.~K.} \bibnamefont{Frame}}, Ph.D. thesis
  (\bibinfo{year}{2019}), \eprint{1905.02782}.

\bibitem[{\citenamefont{König et~al.}(2019)\citenamefont{König, Ekström,
  Hebeler, Lee, and Schwenk}}]{Konig:2019adq}
\bibinfo{author}{\bibfnamefont{S.}~\bibnamefont{König}},
  \bibinfo{author}{\bibfnamefont{A.}~\bibnamefont{Ekström}},
  \bibinfo{author}{\bibfnamefont{K.}~\bibnamefont{Hebeler}},
  \bibinfo{author}{\bibfnamefont{D.}~\bibnamefont{Lee}}, \bibnamefont{and}
  \bibinfo{author}{\bibfnamefont{A.}~\bibnamefont{Schwenk}}
  (\bibinfo{year}{2019}), \eprint{1909.08446}.

\bibitem[{\citenamefont{Ekström and Hagen}(2019)}]{Ekstrom:2019lss}
\bibinfo{author}{\bibfnamefont{A.}~\bibnamefont{Ekström}} \bibnamefont{and}
  \bibinfo{author}{\bibfnamefont{G.}~\bibnamefont{Hagen}},
  \bibinfo{journal}{Phys. Rev. Lett.} \textbf{\bibinfo{volume}{123}},
  \bibinfo{pages}{252501} (\bibinfo{year}{2019}), \eprint{1910.02922}.

\bibitem[{\citenamefont{Demol et~al.}(2020)\citenamefont{Demol, Duguet,
  Ekström, Frosini, Hebeler, König, Lee, Schwenk, Somà, and
  Tichai}}]{Demol:2019yjt}
\bibinfo{author}{\bibfnamefont{P.}~\bibnamefont{Demol}},
  \bibinfo{author}{\bibfnamefont{T.}~\bibnamefont{Duguet}},
  \bibinfo{author}{\bibfnamefont{A.}~\bibnamefont{Ekström}},
  \bibinfo{author}{\bibfnamefont{M.}~\bibnamefont{Frosini}},
  \bibinfo{author}{\bibfnamefont{K.}~\bibnamefont{Hebeler}},
  \bibinfo{author}{\bibfnamefont{S.}~\bibnamefont{König}},
  \bibinfo{author}{\bibfnamefont{D.}~\bibnamefont{Lee}},
  \bibinfo{author}{\bibfnamefont{A.}~\bibnamefont{Schwenk}},
  \bibinfo{author}{\bibfnamefont{V.}~\bibnamefont{Somà}}, \bibnamefont{and}
  \bibinfo{author}{\bibfnamefont{A.}~\bibnamefont{Tichai}},
  \bibinfo{journal}{Phys. Rev.} \textbf{\bibinfo{volume}{C101}},
  \bibinfo{pages}{041302(R)} (\bibinfo{year}{2020}), \eprint{1911.12578}.

\bibitem[{\citenamefont{Ekström et~al.}(2015)\citenamefont{Ekström, Jansen,
  Wendt, Hagen, Papenbrock, Carlsson, Forssén, Hjorth-Jensen, Navrátil, and
  Nazarewicz}}]{Ekstrom:2015rta}
\bibinfo{author}{\bibfnamefont{A.}~\bibnamefont{Ekström}},
  \bibinfo{author}{\bibfnamefont{G.~R.} \bibnamefont{Jansen}},
  \bibinfo{author}{\bibfnamefont{K.~A.} \bibnamefont{Wendt}},
  \bibinfo{author}{\bibfnamefont{G.}~\bibnamefont{Hagen}},
  \bibinfo{author}{\bibfnamefont{T.}~\bibnamefont{Papenbrock}},
  \bibinfo{author}{\bibfnamefont{B.~D.} \bibnamefont{Carlsson}},
  \bibinfo{author}{\bibfnamefont{C.}~\bibnamefont{Forssén}},
  \bibinfo{author}{\bibfnamefont{M.}~\bibnamefont{Hjorth-Jensen}},
  \bibinfo{author}{\bibfnamefont{P.}~\bibnamefont{Navrátil}},
  \bibnamefont{and}
  \bibinfo{author}{\bibfnamefont{W.}~\bibnamefont{Nazarewicz}},
  \bibinfo{journal}{Phys. Rev.} \textbf{\bibinfo{volume}{C91}},
  \bibinfo{pages}{051301} (\bibinfo{year}{2015}), \eprint{1502.04682}.

\bibitem[{\citenamefont{Elhatisari et~al.}(2016)}]{Elhatisari:2016owd}
\bibinfo{author}{\bibfnamefont{S.}~\bibnamefont{Elhatisari}}
  \bibnamefont{et~al.}, \bibinfo{journal}{Phys. Rev. Lett.}
  \textbf{\bibinfo{volume}{117}}, \bibinfo{pages}{132501}
  (\bibinfo{year}{2016}), \eprint{1602.04539}.

\bibitem[{\citenamefont{Lapoux et~al.}(2016)\citenamefont{Lapoux, Somà,
  Barbieri, Hergert, Holt, and Stroberg}}]{Lapoux:2016exf}
\bibinfo{author}{\bibfnamefont{V.}~\bibnamefont{Lapoux}},
  \bibinfo{author}{\bibfnamefont{V.}~\bibnamefont{Somà}},
  \bibinfo{author}{\bibfnamefont{C.}~\bibnamefont{Barbieri}},
  \bibinfo{author}{\bibfnamefont{H.}~\bibnamefont{Hergert}},
  \bibinfo{author}{\bibfnamefont{J.~D.} \bibnamefont{Holt}}, \bibnamefont{and}
  \bibinfo{author}{\bibfnamefont{S.~R.} \bibnamefont{Stroberg}},
  \bibinfo{journal}{Phys. Rev. Lett.} \textbf{\bibinfo{volume}{117}},
  \bibinfo{pages}{052501} (\bibinfo{year}{2016}), \eprint{1605.07885}.

\bibitem[{\citenamefont{Piarulli et~al.}(2018)}]{Piarulli:2017dwd}
\bibinfo{author}{\bibfnamefont{M.}~\bibnamefont{Piarulli}}
  \bibnamefont{et~al.}, \bibinfo{journal}{Phys. Rev. Lett.}
  \textbf{\bibinfo{volume}{120}}, \bibinfo{pages}{052503}
  (\bibinfo{year}{2018}), \eprint{1707.02883}.

\bibitem[{\citenamefont{Lu et~al.}(2019)\citenamefont{Lu, Li, Elhatisari, Lee,
  Epelbaum, and Meißner}}]{Lu:2018bat}
\bibinfo{author}{\bibfnamefont{B.-N.} \bibnamefont{Lu}},
  \bibinfo{author}{\bibfnamefont{N.}~\bibnamefont{Li}},
  \bibinfo{author}{\bibfnamefont{S.}~\bibnamefont{Elhatisari}},
  \bibinfo{author}{\bibfnamefont{D.}~\bibnamefont{Lee}},
  \bibinfo{author}{\bibfnamefont{E.}~\bibnamefont{Epelbaum}}, \bibnamefont{and}
  \bibinfo{author}{\bibfnamefont{U.-G.} \bibnamefont{Meißner}},
  \bibinfo{journal}{Phys. Lett.} \textbf{\bibinfo{volume}{B797}},
  \bibinfo{pages}{134863} (\bibinfo{year}{2019}), \eprint{1812.10928}.

\bibitem[{\citenamefont{Binder et~al.}(2018)}]{Binder:2018pgl}
\bibinfo{author}{\bibfnamefont{S.}~\bibnamefont{Binder}} \bibnamefont{et~al.}
  (\bibinfo{collaboration}{LENPIC}), \bibinfo{journal}{Phys. Rev.}
  \textbf{\bibinfo{volume}{C98}}, \bibinfo{pages}{014002}
  (\bibinfo{year}{2018}), \eprint{1802.08584}.

\bibitem[{\citenamefont{Somà et~al.}(2020)\citenamefont{Somà, Navrátil,
  Raimondi, Barbieri, and Duguet}}]{Soma:2019bso}
\bibinfo{author}{\bibfnamefont{V.}~\bibnamefont{Somà}},
  \bibinfo{author}{\bibfnamefont{P.}~\bibnamefont{Navrátil}},
  \bibinfo{author}{\bibfnamefont{F.}~\bibnamefont{Raimondi}},
  \bibinfo{author}{\bibfnamefont{C.}~\bibnamefont{Barbieri}}, \bibnamefont{and}
  \bibinfo{author}{\bibfnamefont{T.}~\bibnamefont{Duguet}},
  \bibinfo{journal}{Phys. Rev.} \textbf{\bibinfo{volume}{C101}},
  \bibinfo{pages}{014318} (\bibinfo{year}{2020}), \eprint{1907.09790}.

\bibitem[{\citenamefont{Gandolfi et~al.}(2020)\citenamefont{Gandolfi,
  Lonardoni, Lovato, and Piarulli}}]{Gandolfi:2020pbj}
\bibinfo{author}{\bibfnamefont{S.}~\bibnamefont{Gandolfi}},
  \bibinfo{author}{\bibfnamefont{D.}~\bibnamefont{Lonardoni}},
  \bibinfo{author}{\bibfnamefont{A.}~\bibnamefont{Lovato}}, \bibnamefont{and}
  \bibinfo{author}{\bibfnamefont{M.}~\bibnamefont{Piarulli}}
  (\bibinfo{year}{2020}), \eprint{2001.01374}.

\bibitem[{\citenamefont{Tews et~al.}(2020)\citenamefont{Tews, Davoudi,
  Ekström, Holt, and Lynn}}]{Tews:2020hgp}
\bibinfo{author}{\bibfnamefont{I.}~\bibnamefont{Tews}},
  \bibinfo{author}{\bibfnamefont{Z.}~\bibnamefont{Davoudi}},
  \bibinfo{author}{\bibfnamefont{A.}~\bibnamefont{Ekström}},
  \bibinfo{author}{\bibfnamefont{J.~D.} \bibnamefont{Holt}}, \bibnamefont{and}
  \bibinfo{author}{\bibfnamefont{J.~E.} \bibnamefont{Lynn}}
  (\bibinfo{year}{2020}), \eprint{2001.03334}.

\bibitem[{\citenamefont{Leggett}(1980)}]{Leggett:1980pro}
\bibinfo{author}{\bibfnamefont{A.~J.} \bibnamefont{Leggett}}, in
  \emph{\bibinfo{booktitle}{Modern Trends in the Theory of Condensed Matter.
  Proceedings of the XVIth Karpacz Winter School of Theoretical Physics,
  Karpacz, Poland, 1980}} (\bibinfo{publisher}{Springer-Verlag, Berlin},
  \bibinfo{year}{1980}), p.~\bibinfo{pages}{13}.

\bibitem[{\citenamefont{Nozieres and Schmitt-Rink}(1985)}]{Nozieres:1985JLTP}
\bibinfo{author}{\bibfnamefont{P.}~\bibnamefont{Nozieres}} \bibnamefont{and}
  \bibinfo{author}{\bibfnamefont{S.}~\bibnamefont{Schmitt-Rink}},
  \bibinfo{journal}{J. Low Temp. Phys.} \textbf{\bibinfo{volume}{59}},
  \bibinfo{pages}{195} (\bibinfo{year}{1985}).

\bibitem[{\citenamefont{O'Hara et~al.}(2002)\citenamefont{O'Hara, Hemmer, Gehm,
  Granade, and Thomas}}]{OHara:2002}
\bibinfo{author}{\bibfnamefont{K.~M.} \bibnamefont{O'Hara}},
  \bibinfo{author}{\bibfnamefont{S.~L.} \bibnamefont{Hemmer}},
  \bibinfo{author}{\bibfnamefont{M.~E.} \bibnamefont{Gehm}},
  \bibinfo{author}{\bibfnamefont{S.~R.} \bibnamefont{Granade}},
  \bibnamefont{and} \bibinfo{author}{\bibfnamefont{J.~E.}
  \bibnamefont{Thomas}}, \bibinfo{journal}{Science}
  \textbf{\bibinfo{volume}{298}}, \bibinfo{pages}{2179} (\bibinfo{year}{2002}).

\bibitem[{\citenamefont{Gupta et~al.}(2003)\citenamefont{Gupta, Hadzibabic,
  Zwierlein, Stan, Dieckmann, Schunck, van Kempen, Verhaar, and
  Ketterle}}]{Gupta:2002}
\bibinfo{author}{\bibfnamefont{S.}~\bibnamefont{Gupta}},
  \bibinfo{author}{\bibfnamefont{Z.}~\bibnamefont{Hadzibabic}},
  \bibinfo{author}{\bibfnamefont{M.~W.} \bibnamefont{Zwierlein}},
  \bibinfo{author}{\bibfnamefont{C.~A.} \bibnamefont{Stan}},
  \bibinfo{author}{\bibfnamefont{K.}~\bibnamefont{Dieckmann}},
  \bibinfo{author}{\bibfnamefont{C.~H.} \bibnamefont{Schunck}},
  \bibinfo{author}{\bibfnamefont{E.~G.~M.} \bibnamefont{van Kempen}},
  \bibinfo{author}{\bibfnamefont{B.~J.} \bibnamefont{Verhaar}},
  \bibnamefont{and} \bibinfo{author}{\bibfnamefont{W.}~\bibnamefont{Ketterle}},
  \bibinfo{journal}{Science} \textbf{\bibinfo{volume}{300}},
  \bibinfo{pages}{1723} (\bibinfo{year}{2003}).

\bibitem[{\citenamefont{Regal and Jin}(2003)}]{Regal:2003}
\bibinfo{author}{\bibfnamefont{C.~A.} \bibnamefont{Regal}} \bibnamefont{and}
  \bibinfo{author}{\bibfnamefont{D.~S.} \bibnamefont{Jin}},
  \bibinfo{journal}{Phys. Rev. Lett.} \textbf{\bibinfo{volume}{90}},
  \bibinfo{pages}{230404} (\bibinfo{year}{2003}).

\bibitem[{\citenamefont{{Ku} et~al.}(2012)\citenamefont{{Ku}, {Sommer},
  {Cheuk}, and {Zwierlein}}}]{Ku:2012a}
\bibinfo{author}{\bibfnamefont{M.~J.~H.} \bibnamefont{{Ku}}},
  \bibinfo{author}{\bibfnamefont{A.~T.} \bibnamefont{{Sommer}}},
  \bibinfo{author}{\bibfnamefont{L.~W.} \bibnamefont{{Cheuk}}},
  \bibnamefont{and} \bibinfo{author}{\bibfnamefont{M.~W.}
  \bibnamefont{{Zwierlein}}}, \bibinfo{journal}{Science}
  \textbf{\bibinfo{volume}{335}}, \bibinfo{pages}{563} (\bibinfo{year}{2012}),
  \eprint{1110.3309}.

\bibitem[{\citenamefont{{Bour} et~al.}(2011)\citenamefont{{Bour}, {Li}, {Lee},
  {Mei{\ss}ner}, and {Mitas}}}]{Bour:2011ad}
\bibinfo{author}{\bibfnamefont{S.}~\bibnamefont{{Bour}}},
  \bibinfo{author}{\bibfnamefont{X.}~\bibnamefont{{Li}}},
  \bibinfo{author}{\bibfnamefont{D.}~\bibnamefont{{Lee}}},
  \bibinfo{author}{\bibfnamefont{U.-G.} \bibnamefont{{Mei{\ss}ner}}},
  \bibnamefont{and} \bibinfo{author}{\bibfnamefont{L.}~\bibnamefont{{Mitas}}},
  \bibinfo{journal}{Phys. Rev.} \textbf{\bibinfo{volume}{A83}},
  \bibinfo{pages}{063619} (\bibinfo{year}{2011}), \eprint{arXiv:1104.2102
  [cond-mat.quant-gas]}.

\bibitem[{\citenamefont{Kompaniets}(2016)}]{Kompaniets:2016lmy}
\bibinfo{author}{\bibfnamefont{M.}~\bibnamefont{Kompaniets}},
  \bibinfo{journal}{J. Phys. Conf. Ser.} \textbf{\bibinfo{volume}{762}},
  \bibinfo{pages}{012075} (\bibinfo{year}{2016}), \eprint{1604.04108}.

\bibitem[{\citenamefont{{Van Houcke} et~al.}(2020)\citenamefont{{Van Houcke},
  {Werner}, and {Rossi}}}]{VanHoucke:2020}
\bibinfo{author}{\bibfnamefont{K.}~\bibnamefont{{Van Houcke}}},
  \bibinfo{author}{\bibfnamefont{F.}~\bibnamefont{{Werner}}}, \bibnamefont{and}
  \bibinfo{author}{\bibfnamefont{R.}~\bibnamefont{{Rossi}}},
  \bibinfo{journal}{\prb} \textbf{\bibinfo{volume}{101}}, \bibinfo{eid}{045134}
  (\bibinfo{year}{2020}), \eprint{1911.01345}.

\bibitem[{\citenamefont{Bender and Wu}(1969)}]{Bender:1969si}
\bibinfo{author}{\bibfnamefont{C.~M.} \bibnamefont{Bender}} \bibnamefont{and}
  \bibinfo{author}{\bibfnamefont{T.~T.} \bibnamefont{Wu}},
  \bibinfo{journal}{Phys. Rev.} \textbf{\bibinfo{volume}{184}},
  \bibinfo{pages}{1231} (\bibinfo{year}{1969}).

\bibitem[{\citenamefont{Parwani}(2001)}]{Parwani:2000rr}
\bibinfo{author}{\bibfnamefont{R.~R.} \bibnamefont{Parwani}},
  \bibinfo{journal}{Phys. Rev.} \textbf{\bibinfo{volume}{D63}},
  \bibinfo{pages}{054014} (\bibinfo{year}{2001}), \eprint{hep-ph/0010234}.

\bibitem[{\citenamefont{Hamprecht and Kleinert}(2003)}]{Hamprecht:2003vr}
\bibinfo{author}{\bibfnamefont{B.}~\bibnamefont{Hamprecht}} \bibnamefont{and}
  \bibinfo{author}{\bibfnamefont{H.}~\bibnamefont{Kleinert}},
  \bibinfo{journal}{Phys. Lett.} \textbf{\bibinfo{volume}{B564}},
  \bibinfo{pages}{111} (\bibinfo{year}{2003}), \eprint{hep-th/0302124}.

\bibitem[{\citenamefont{Argyres and Unsal}(2012)}]{Argyres:2012ka}
\bibinfo{author}{\bibfnamefont{P.~C.} \bibnamefont{Argyres}} \bibnamefont{and}
  \bibinfo{author}{\bibfnamefont{M.}~\bibnamefont{Unsal}},
  \bibinfo{journal}{JHEP} \textbf{\bibinfo{volume}{08}}, \bibinfo{pages}{063}
  (\bibinfo{year}{2012}), \eprint{1206.1890}.

\bibitem[{\citenamefont{Cherman et~al.}(2015)\citenamefont{Cherman, Dorigoni,
  and Unsal}}]{Cherman:2014ofa}
\bibinfo{author}{\bibfnamefont{A.}~\bibnamefont{Cherman}},
  \bibinfo{author}{\bibfnamefont{D.}~\bibnamefont{Dorigoni}}, \bibnamefont{and}
  \bibinfo{author}{\bibfnamefont{M.}~\bibnamefont{Unsal}},
  \bibinfo{journal}{JHEP} \textbf{\bibinfo{volume}{10}}, \bibinfo{pages}{056}
  (\bibinfo{year}{2015}), \eprint{1403.1277}.

\end{thebibliography}
\bibliographystyle{apsrev}
       
\beginsupplement
     
 \clearpage     
        
\section{Supplemental Materials}

\subsection{Details of the Hamiltonians used}
In the main text we consider several examples of one parameter families of Hamiltonians, $H(c) = H_0 + cH_1$, where $H_0$ and $H_1$ are Hermitian matrices.  These are labelled as Model 1A, 1B, 1C, 2, and 3.  We describe each of the models in detail.

\subsubsection{Model 1}
In Models 1A, 1B and 1C, we take $H_0$ to be diagonal matrices, while $H_1$ has both diagonal and off-diagonal elements. For all three model calculations, the target value is $c_t = 0.2$.  In Model 1A, we take $H_0$ to be the matrix with elements $H_0(n,n) = n$  for $n=1,\cdots,800$. We choose $H_1$ to be $H_1(n,n) = n$ for $n=1,\cdots,800$ and 
\begin{equation}
 H_1(n+2,n) = H_1(n,n+2) = 1,
\end{equation}
for $n=1,\cdots,798$. For this model, the nearest branch point to the origin is located at $c=-0.559\pm0.497i$. 

In Model 1B, we take $H_0$ to be the matrix with elements $H_0(n,n) = 2n$  for $n=1,\cdots,800$. We choose $H_1$ to be $H_1(n,n) = n$ for $n=1,\cdots,800$ and
\begin{equation}
H_1(n+2,n) = H_1(n,n+2) = 1,
\end{equation}
for $n=1,\cdots,798$. For this model, the nearest branch point to the origin is located at $c=-1.422\pm 0.503i$. 

In Model 1C, we take $H_0$ to be the matrix with elements $H_0(n,n) = 100n$  for $n=1,\cdots,800$. We choose $H_1$ to be $H_1(1,1) = -100$, $H_1(2,2) = -200$, and take $H_1(n,n) = -75n$ for $n=3,\cdots,800$.  For the off-diagonal entries we set
\begin{equation}
H_1(n+1,n) = H_1(n,n+1) = 1,
\end{equation}
for $n=1,\cdots,799$. For this model, the nearest branch point to the origin is located at
$c=0.907\pm 0.255i$. 

\subsubsection{Model 2}

As described in the main text, Model 2 corresponds to a system of two-component fermions with zero range interactions in three dimensions. At weak coupling the many-body system is a BCS superfluid, while at strong coupling it is a Bose-Einstein condensate.  The crossover region between these two phases contains the unitary limit where the scattering length diverges.  We consider the lattice Hamiltonian for two spin-up fermions and two spin-down fermions in a cubic periodic box as described in Ref.~\cite{Bour:2011ad}.  The corresponding linear space has $4^9$ = 262144 dimensions, and we use projection operators to remove all unphysical states without the proper antisymmetrization.

Let $\Vec{n}$ denote the spatial lattice points on a three dimensional $L^3$ periodic lattice. We
use lattice units where physical quantities are multiplied by powers of the spatial lattice
spacing to make the combination dimensionless. Let the lattice annihilation operators for the spin-up and spin-down fermions be $a_{\uparrow}(\vec{n})$ and $a_{\downarrow}(\vec{n})$ respectively.  We use the free non-relativistic lattice Hamiltonian
\begin{align}
    & H_{\text{free}} = \frac{3}{m}\sum_{i=\uparrow,\downarrow}\sum_{\Vec{n}}
    a_{i}^\dagger(\Vec{n})a_{i}(\Vec{n}) \nonumber \\
    -&\frac{1}{2m}\sum_{l=1,2,3}
    \sum_{i=\uparrow,\downarrow}\sum_{\Vec{n}}
    a_{i}^\dagger(\Vec{n})\Big[a_{i}(\Vec{n}+\hat{l})+a_{i}(\Vec{n}-\hat{l})\Big].
\end{align}
We add to the free Hamiltonian a single-site contact interaction,
\begin{equation}
    H_{\text{free}} + C\sum_{\Vec{n}}\rho_{\uparrow}(\Vec{n})\rho_{\downarrow}(\Vec{n}),
\end{equation}
where
\begin{align}
    \rho_{\uparrow}(\Vec{n}) &= a_{\uparrow}^\dagger(\Vec{n})a_{\uparrow}(\Vec{n}), \\
    \rho_{\downarrow}(\Vec{n}) &= a_{\downarrow}^\dagger(\Vec{n})a_{\downarrow}(\Vec{n}).
\end{align}
Our control parameter $c$ corresponds to the product of the particle mass $m$ and interaction coupling $C$.  The parameter value $c=-3.957$ corresponds with the unitary limit. Since the point $c = 0$ corresponds to a non-interacting system with special symmetries and degeneracies, we choose the training vectors at a more general point on the weak-coupling BCS side at $c = -0.4695$.

\subsubsection{Model 3}

In Model 3 we take $H_0$ to be a $500 \times 500$ diagonal matrix with entries $H_0(n,n) = 100n$ for $n=1,\cdots,500$.  $H_1$ is a $500 \times 500$ matrix with nonzero entries as follows: 
\begin{align}
H_1(1,1) &= 40, \\
H_1(2,2) &= -80, \\
H_1(3,3) &= -180, \\
H_1(4,4) &= -260, \\
H_1(5,5) &= -320, \\
H_1(6,6) &= -335,
\end{align}
for $n =  1,\cdots, 499$:
\begin{equation}
H_1(n+1,n) = H_1(n,n+1)=2,
\end{equation}
for $n =  1,\cdots, 498$:
\begin{equation}
H_1(n+2,n) = H_1(n,n+2)=5,
\end{equation}
for $n =  1,\cdots, 497$:
\begin{equation}
H_1(n+3,n) = H_1(n,n+3)=5,
\end{equation}
for $n =  7,\cdots, 500$:
\begin{equation}
H_1(n,n) = 50n.
\end{equation}

    \begin{figure}
                \centering
                \includegraphics[width=8.4cm]{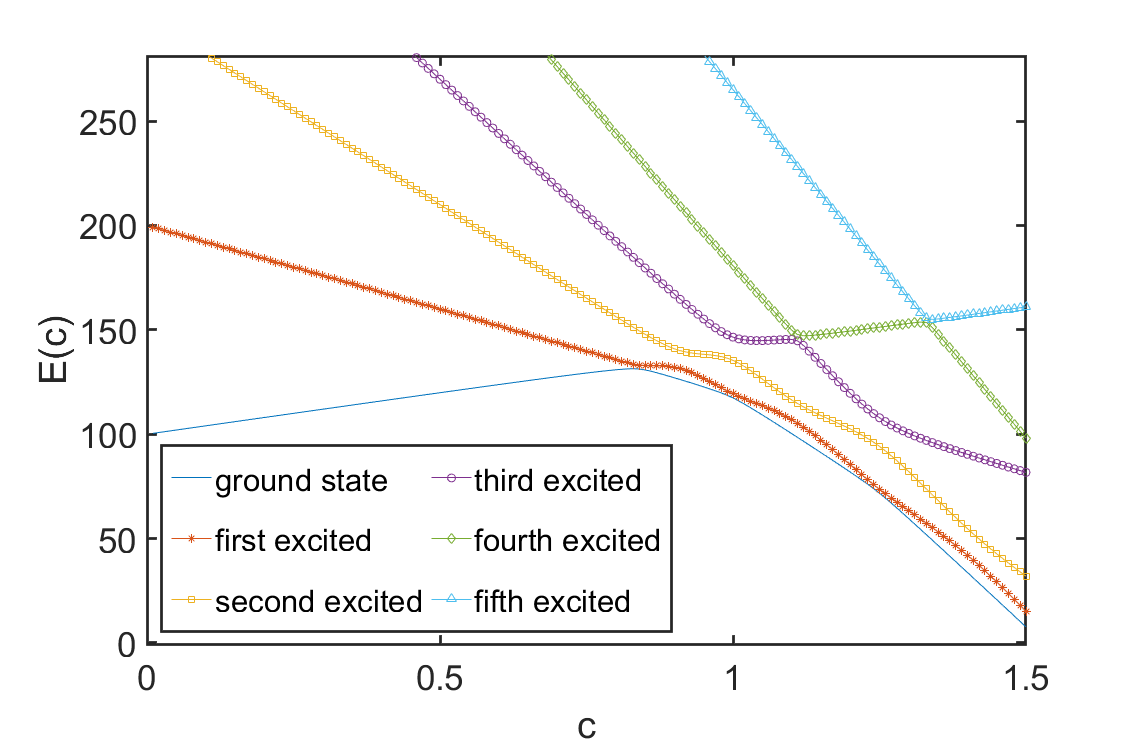}\\
                \caption{The lowest six energies of Model 3 as a function of $c$.}
                \label{level_crossing}
        \end{figure}

Model 3 is chosen such that perturbation  theory  will  break  down  due  to  several sharp avoided level crossings. In Fig.~\ref{level_crossing} we show the energies of the lowest six energies as a function of the control parameter $c$.  The closest branch point to $c=0$ occurs very close to the real axis near $c=0.84$.  The first avoided level crossing near $c=0.84$ can be seen in the figure.

\begin{figure}
        \centering
        \includegraphics[width=8.4cm]{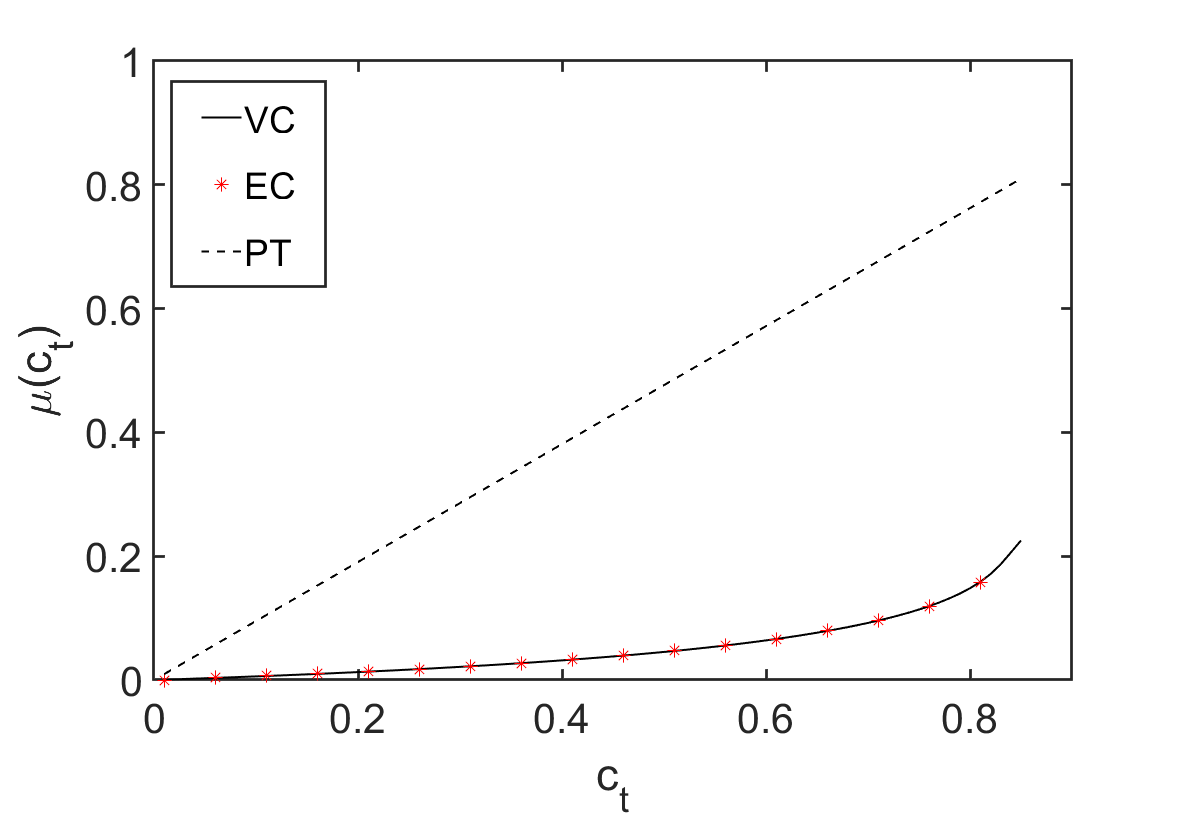}\\
        \caption{Comparison of the convergence ratios $\mu^{\rm VC}$, $\mu^{\rm EC}$, and $\mu^{\rm PT}$  versus $c_t$ for Model 3 with $N = 20$ and $N' = 0$.} 
        \label{PT_VC}
\end{figure}

\begin{figure}
        \centering
        \includegraphics[width=8.4cm]{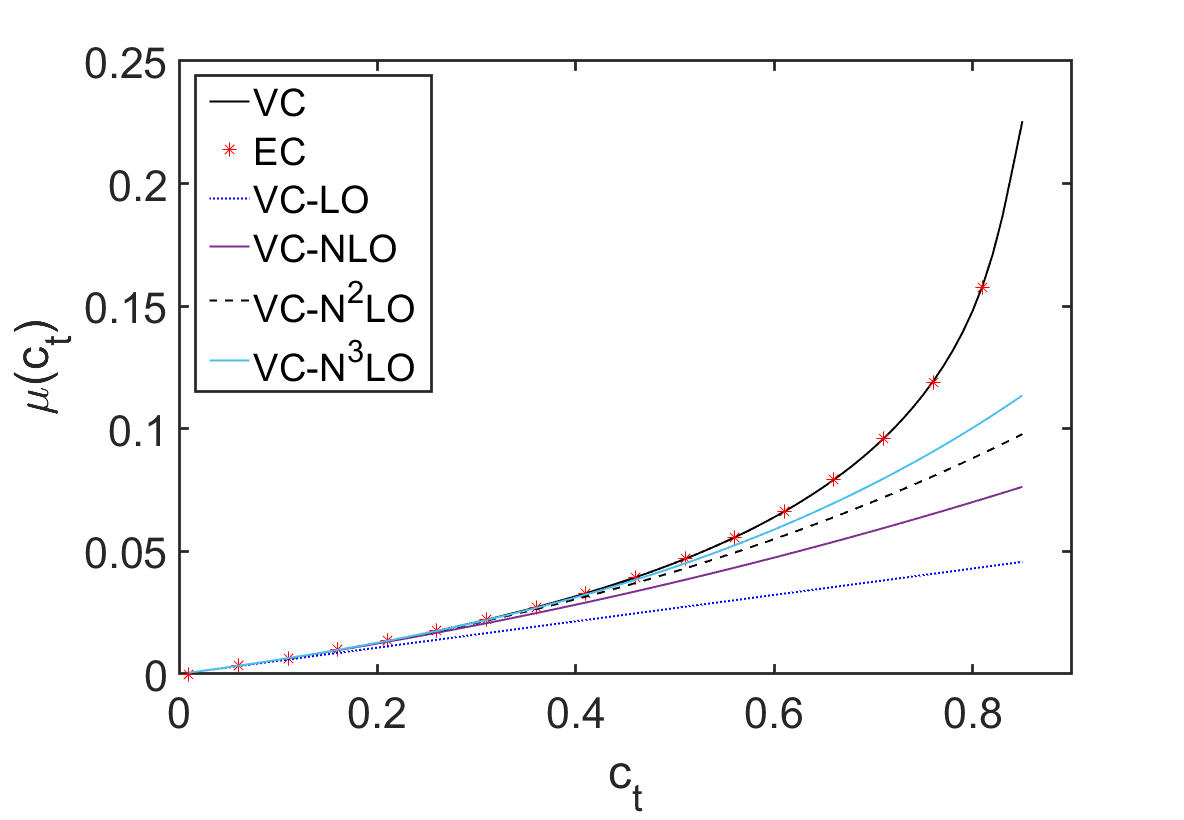}\\
        \caption{Plots of the convergence ratios $\mu^{\rm VC}$, $\mu^{\rm EC}$, and the LO, NLO, N$^2$LO, and N$^3$LO approximations to $\mu^{\rm VC}$ versus $c_t$ for Model 3 with $N = 20$ and $N' = 0$.}
        \label{PT_VC_VC-PT}
\end{figure}  

In Fig.~\ref{PT_VC} we show a comparison of the convergence ratios $\mu^{\rm VC}$, $\mu^{\rm EC}$, and $\mu^{\rm PT}$ versus $c_t$ for $N = 20$ and $N' = 0$, for Model 3.  In the limit $|N-N'|\rightarrow \infty$, the convergence ratio $\mu^{\rm PT}$ will cross the value $1$ for $c_t$ near 0.84, signalling the divergence of perturbation theory. As Fig.~\ref{PT_VC} shows, $\mu^{\rm PT}$ is approaching 1 near $c_t = 0.84$ for $N = 20$ and $N' = 0$.  On the other hand, $\mu^{\rm VC}$ and $\mu^{\rm EC}$ are in close agreement with each other and remain well below $1$ near $c_t = 0.84$.

In Fig.~\ref{PT_VC_VC-PT} we plot $\mu^{\rm VC}$ and the LO, NLO, N$^2$LO, and N$^3$LO approximations to $\mu^{\rm VC}$ for $N = 20$ and $N' = 0$.  Once again we see that the series expansion in Eq.~(\ref{VC-PT}) converges for $c_t$ within the radius of convergence of perturbation theory, which for this example corresponds to $c = 0.84$.  

\subsection{Convergence ratio for different values of $N$}
\begin{figure}
        \centering
        \includegraphics[width=8.4cm]{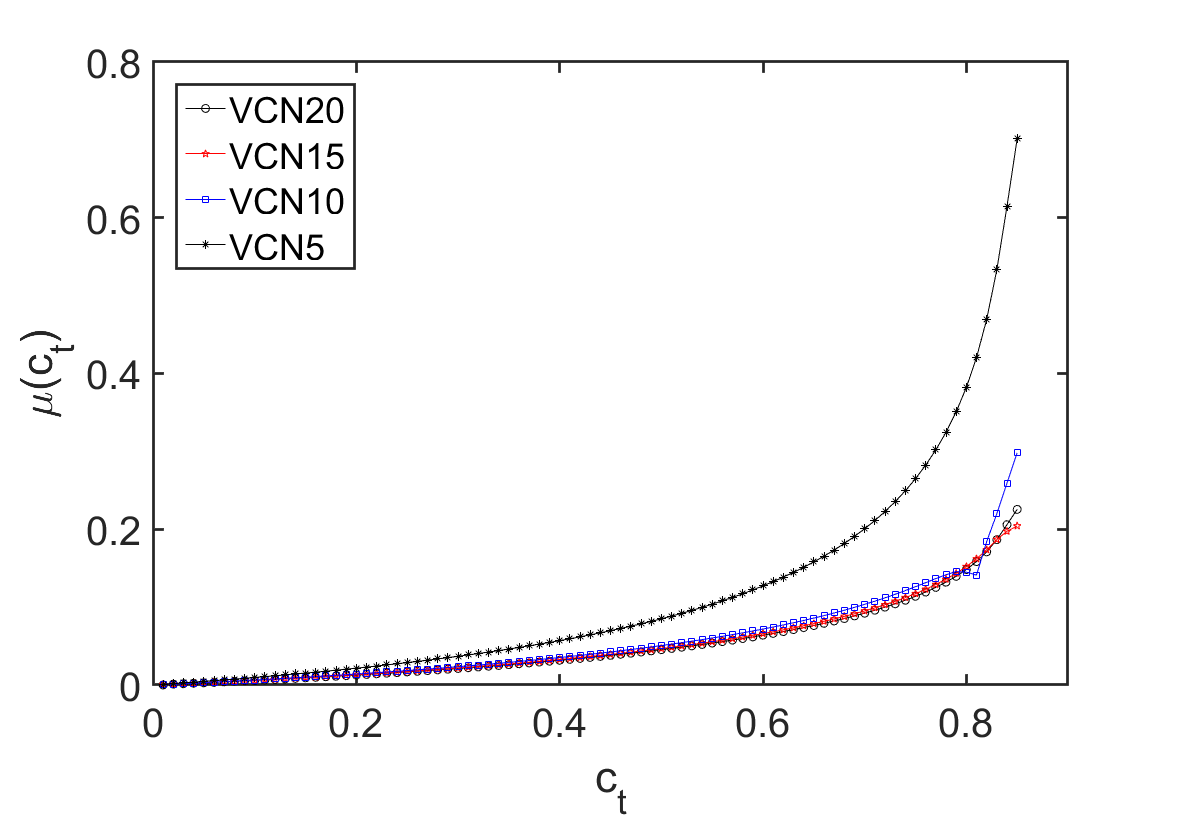}\\
        \caption{Plots of the convergence ratio $\mu^{\rm VC}$  versus $c_t$ for Model 3 with $N' = 0$ and $N = 5,10,15,20$.}
        \label{VC_N}
\end{figure}  
The convergence ratio $\mu^{\rm VC}$ provides information about the rate of convergence of vector continuation in the limit of many training vectors.  In Fig.~\ref{VC_N} we plot $\mu^{\rm VC}$ for Model 3 for $N'=0$ and $N = 5,10,15,20$.  We see that indeed the convergence ratio is approaching a common ratio in the limit of large $N$.

\subsection{Convergence of multi-parameter eigenvector continuation}

In the main text we discussed the convergence of eigenvector continuation for the multi-parameter case with $D>1$ dimensions.  For the problem of convergence for all points at some distance from the training limit point, we need to include all $(k+D-1)!/[k!(D-1)!]$ partial derivatives at order $k$.   Summing over all possible values for $k$, we conclude that a $D$ parameter calculation with $N_D+1$ training vectors is equivalent to a one parameter calculation with $N_1+1$ training vectors when 
\begin{equation}
    N_D = \frac{(N_1+D)!}{N_1!D!}-1.
\end{equation}

\begin{figure}
        \centering
        \includegraphics[width=8.4cm]{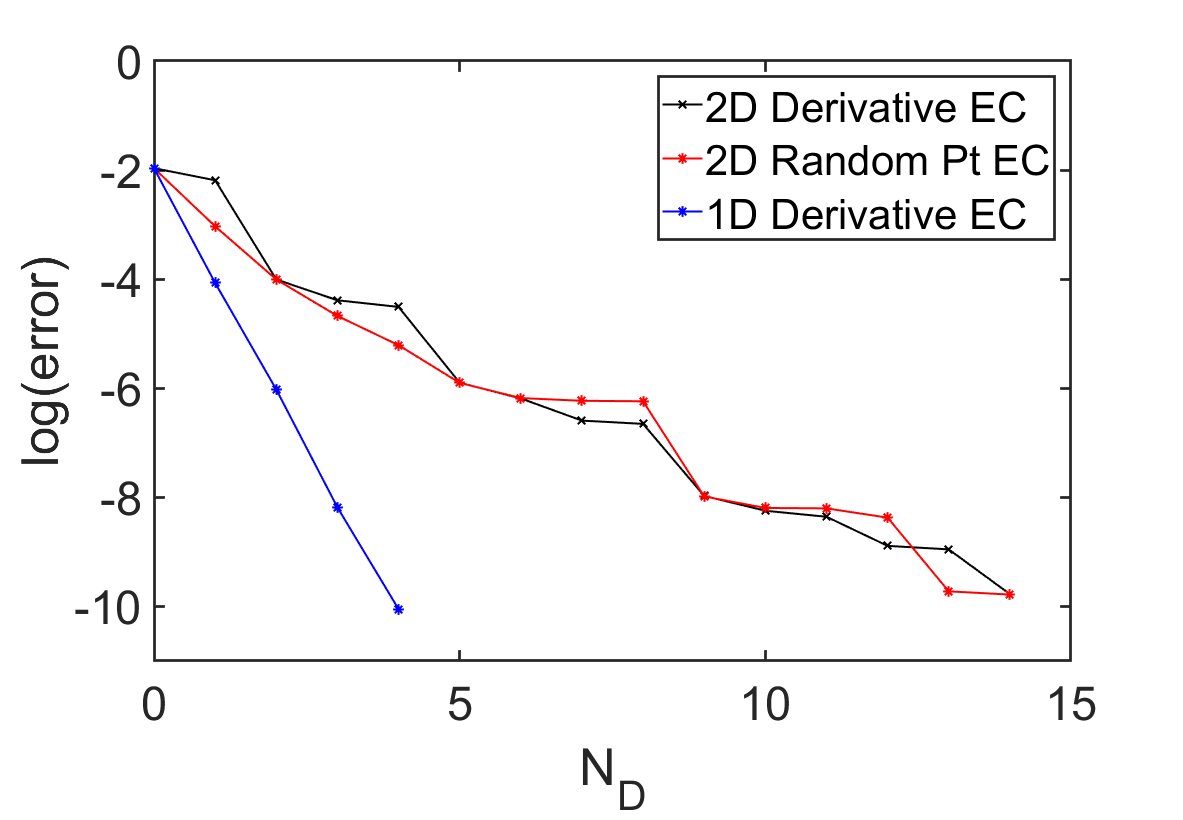}\\
        \caption{Comparison of the logarithm of the EC error versus $N_D$ for $D=1$ and $D=2$ dimensions.} 
        \label{2Dvs1D}
\end{figure}

In order to test the multi-parameter convergence of eigenvector continuation, we consider some numerical examples.  Let $H_0$ be a diagonal matrix with elements $H_0(n,n) = n^{0.1}$ with $n=1,\cdots,500$.  Let $H_1$, $H_2$, and $H_3$ be three different random $500 \times 500$ matrices with each matrix element sampled from a normal distribution with zero mean and standard deviation $10^{-3}$.  We will refer to this random matrix example as Model 4.   
We first consider the convergence of eigenvector continuation for the two parameter Hamiltonian 
\begin{align}
H(c_1,c_2)=H_0 + c_1H_1 +c_2H_2.
\end{align}
We take the training points to be in the neighborhood of the origin $\vec{c}=\vec{0}$. For the target point we pick a random point $\vec{c}_t = (-1.636, 1.150)$.   As shown in Fig.~\ref{2Dvs1D}, we first consider the one parameter case labelled as 1D Derivative EC.  This corresponds to taking directional derivatives 
$[\vec{c}_t \cdot \vec{\nabla}]^k$ acting on $\ket{v(\vec{c})}$ at $\vec{c}=\vec{0}$ for $k=0,\cdots,N_1$, for a total $N_1+1$ training vectors.  We plot the logarithm of the error versus $N_1$.  

For comparison, we also present eigenvector continuation results using $N_2+1$ training vectors for the two dimensional case.  These training vectors correspond to the $N_2+1$ lowest-order partial derivatives $\nabla_1^{k_1}\nabla_2^{k_2}$ acting on $\ket{v(\vec{c})}$ at $\vec{c}=\vec{0}$.  This is labelled as 2D Derivative EC in Fig.~\ref{2Dvs1D}. If we use the error equivalence formula $N_2 = (N_1+2)!/(N_1!2!)-1$, we get
\begin{align}
    & N_1 = 0 \rightarrow N_2 = 0, \nonumber \\
    & N_1 = 1 \rightarrow N_2 = 2, \nonumber \\
    & N_1 = 2 \rightarrow N_2 = 5,  \nonumber \\
    & N_1 = 3 \rightarrow N_2 = 9,  \nonumber \\
    & N_1 = 4 \rightarrow N_2 = 14,  \nonumber \\   
    & N_1 = 5 \rightarrow N_2 = 20.  
\end{align}
As we see in Fig.~\ref{2Dvs1D}, these predictions work quite well.  The errors for $N_1$ are approximately equal to the errors for the corresponding value of $N_2$.  We also show the results we obtain when, instead of using partial derivatives, we simply take training vectors at $N_2+1$ random points in the neighborhood of the origin.  This is labelled as 2D Random EC.  As we can see, this data agrees well with the 2D Derivative EC data.  This is just as one would expect since the partial derivatives are well approximated by finite differences of $\ket{v(\vec{c})}$ at points in the neighborhood of the origin. 

\begin{figure}
        \centering
        \includegraphics[width=8.4cm]{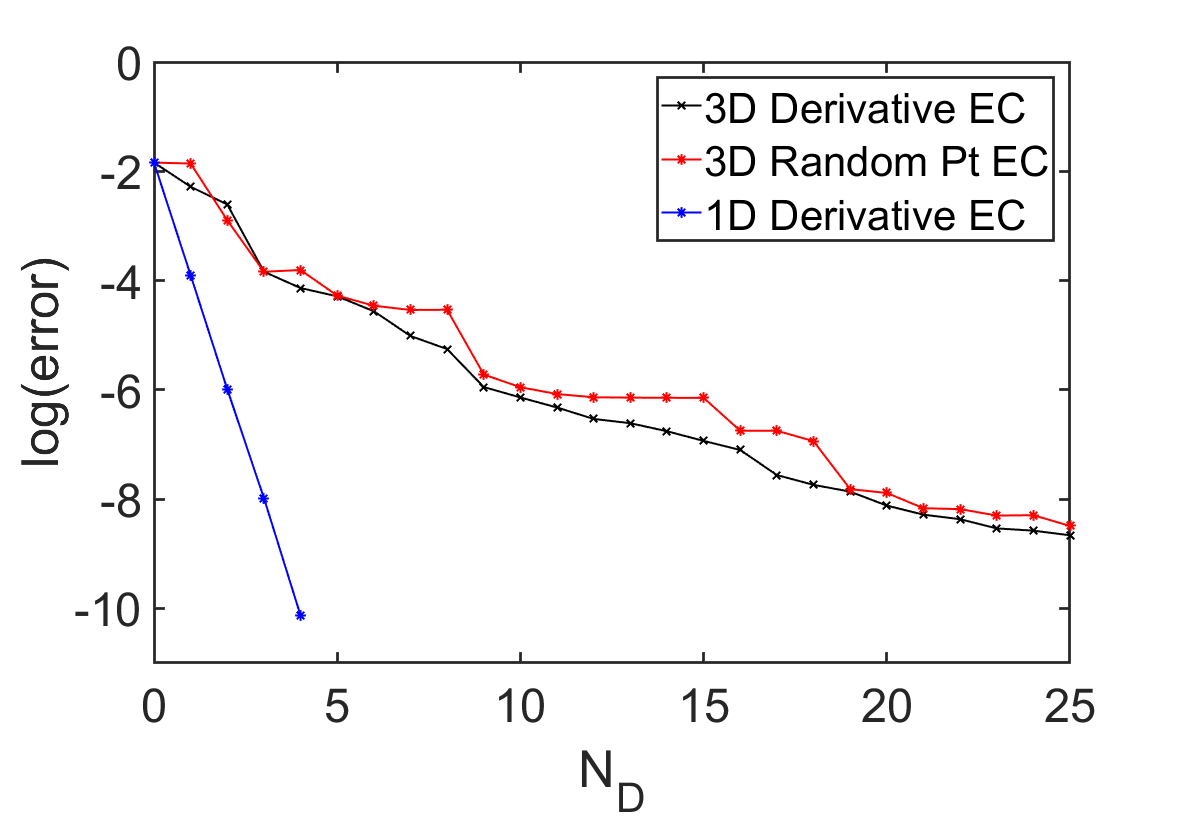}\\
        \caption{Comparison of the logarithm of the EC error versus $N_D$ for $D=1$ and $D=3$ dimensions.} 
        \label{3Dvs1D}
\end{figure}

Using Model 4 again, we now consider the convergence of eigenvector continuation for the three parameter Hamiltonian 
\begin{align}
H(c_1,c_2,c_3)=H_0 + c_1H_1 +c_2H_2+c_3H_3.
\end{align}
with random target point $\vec{c}_t = (-1.034,-1.065,1.341)$.  In this case we have the error equivalence formula $N_3 = (N_1+3)!/(N_1!3!)-1$ and therefore
\begin{align}
    & N_1 = 0 \rightarrow N_3 = 0, \nonumber \\
    & N_1 = 1 \rightarrow N_3 = 3, \nonumber \\
    & N_1 = 2 \rightarrow N_3 = 9,  \nonumber \\
    & N_1 = 3 \rightarrow N_3 = 19.
\end{align}
As we see in Fig.~\ref{3Dvs1D}, these predictions work quite well also.  The errors for $N_1$, labelled as 1D Derivative EC, are approximately equal to the errors for the corresponding value of $N_3$, labelled as 3D Derivative EC.  Once again, we show the results we obtain when we take training vectors at $N_3+1$ random points in the neighborhood of the origin.  This is labelled as 3D Random EC, and the errors match quite well with the 3D Derivative EC results.  

\subsection{Dependence on the positions of training points}
As noted in the main text, the general classification of eigenvector continuation for all possible sets of training points is too large a task for this work or perhaps any single individual study. Nevertheless we can discuss at least one interesting case that illustrates some universal features. As we already noted, the multi-parameter case reduces to the one parameter case if we select training points along a straight line passing through the target point.  

\begin{figure}
        \centering
        \includegraphics[width=8.4cm]{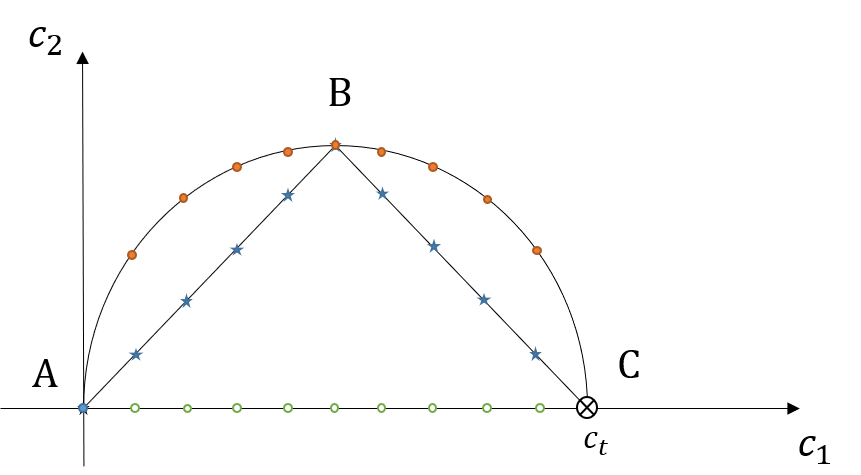}\\
        \caption{We show different choices of path for the EC training vectors. The direct straight line is in green, semicircle is in orange, isosceles right triangle is in blue.} 
        \label{Training_pts_path_CircleTriangleLine}
\end{figure}

\begin{figure}
        \centering
        \includegraphics[width=8.4cm]{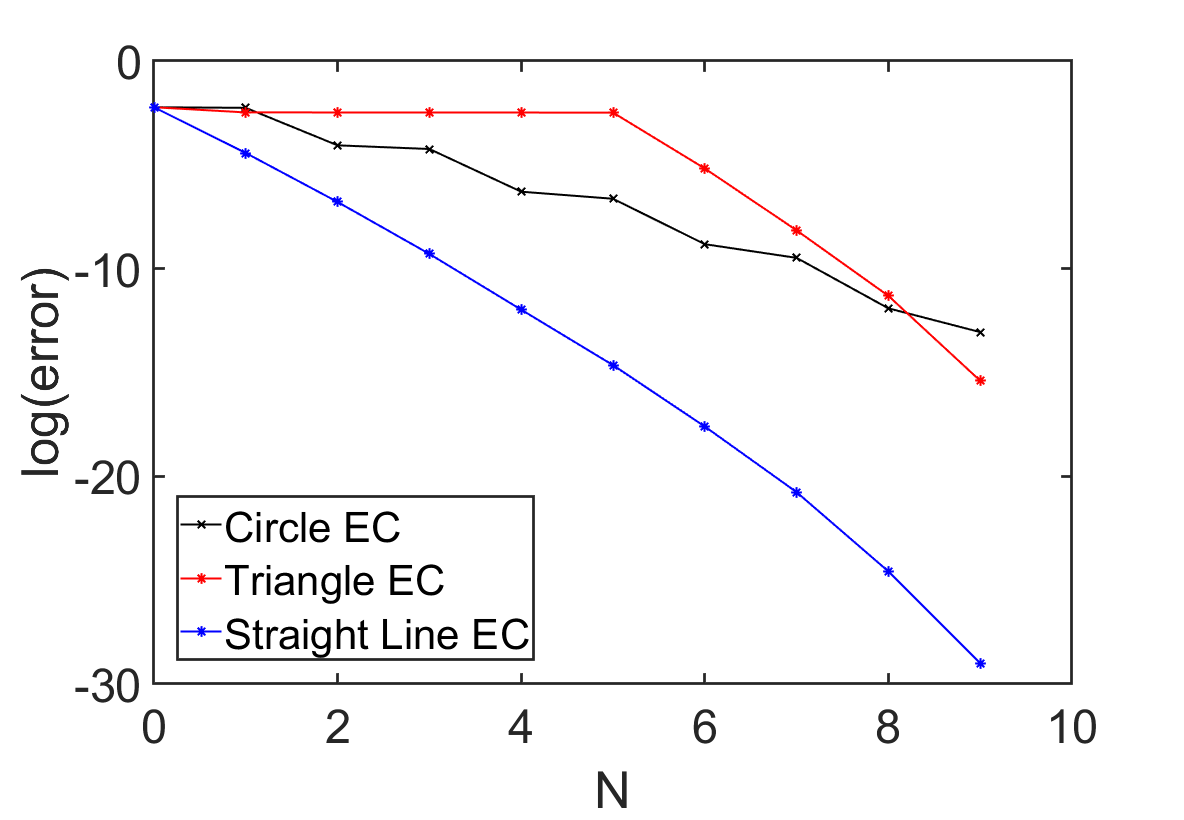}\\
        \caption{Comparison of the logarithm of the error versus order $N$ for different paths for the EC training vectors.  The number of training points corresponds to $N+1$.} 
        \label{CircleTriangleLine}
\end{figure}

If the training points lie along a smooth curved path that passes through the target point, then we expect convergence faster than the general case but slower than the straight line example, due to the curvature of the path and the increased path length.  We illustrate this with the two parameter example for Model 4,
\begin{align}
H(c_1,c_2)=H_0 + c_1H_1 +c_2H_2.
\end{align}
We take the target point, $C$, to be located at $\vec{c}_t=(2,0)$.  We first consider training points evenly spaced along a straight line from the origin, which we label as point $A$, to the target point $C$.  This is shown in Fig.~\ref{Training_pts_path_CircleTriangleLine}. In Fig.~\ref{CircleTriangleLine} we plot the logarithm of the error versus $N$, where the number of training points is given by $N+1$.  We now consider training points along a semicircular arc $ABC$, as shown in Fig.~\ref{Training_pts_path_CircleTriangleLine}. The convergence for this case is displayed in Fig.~\ref{CircleTriangleLine}.  The convergence is slower than for the straight line example but faster than the general two parameter convergence that we saw in Fig.~\ref{2Dvs1D}.  This is in agreement with what we predicted.

Next we consider points along a right isosceles triangle $ABC$ as shown in Fig.~\ref{Training_pts_path_CircleTriangleLine}.  In this example the path has a discontinuity at $B$.  As one might expect, there is almost no reduction in error as we take training points along the line segment $AB$, which does not pass through the target point.  After we reach $B$ and turn the corner to $C$, we see that the convergence rate is similar to the one parameter case again.  This shows that the existence of a smooth path connecting the training points and the target is very important for the convergence rate of eigenvector continuation.

It is also worth mentioning how our example with equally spaced points on the straight line compares with our analysis using derivative vectors.  If the training points are not too far apart, then the convergence is similar to that obtained by replacing the training vectors with $N+1$ derivative vectors evaluated at the centroid of the training data.  If the training points are spaced far apart, however, then the training points far away from the target point are mostly unhelpful, and the convergence is controlled by the training points closest to the target point.  One should then apply the same construction using derivative vectors at the centroid of this smaller subset of training points.

\end{document}